# The quantum-classical boundary and the moments of inertia of physical objects

C. L. Herzenberg

**Abstract**
During the last few years, several studies have proposed the existence of a threshold separating classical from quantum behavior of objects that is dependent on the size and mass of an object as well as being dependent on certain properties usually associated with the universe as a whole. Here, we reexamine the results of these studies and recast the threshold criteria in terms of a critical threshold value for the moments of inertia of physical objects. Physical objects having moments of inertia above this critical threshold value would be expected to behave necessarily in a classical manner in terms of their center of mass motion as entire objects, while physical objects having moments of inertia lower than this threshold value could exhibit quantum behavior unless brought into classicality by other effects. A comparison with observed values of moments of inertia is presented, and the moment of inertia is suggested as a classifying parameter for examination of the quantum versus classical behavior of objects in the mesoscale domain.

**Introduction**

The transition between quantum physics and classical physics has presented a major intellectual enigma for about a century, and has been examined in a variety of experimental and theoretical studies. Interest in various aspects of this topic has continued in recent years (Zurek, 2002; Leggett, 2002; Bhattacharya et al., 2004; Adler and Bassi, 2009). Achieving a clearer understanding of the transition and of the boundary (or, more accurately, the boundary region) separating the behavior of macroscopic bulk matter objects governed by classical physics and the behavior of microscale objects would seem to be important for a number of fields (Habib).

 At present, the transition from quantum to classical physics is widely recognized as having interaction with the environmnet as an essential element, and this is regarded as a fundamental physical origin of classicality (Joos, 2002; Zurek, 2002; Bhattacharya et al., 2002). One aspect of this, quantum decoherence is the dynamical suppression of quantum interference effects induced by the environment, and the key idea promoted by decoherence is the insight that realistic quantum systems are never isolated, but are immersed in the surrounding environment and interact continuously and unavoidably with it (Habib, Schlosshauer). Thus, macroscopic systems never cease to interact uncontrollably with their environment (even including ambient gravitational fields), thus giving rise to decoherence, and hence to their classical behavior (Zeh, 2006; Zeh, website; Penrose; Wang et al.). In short, decoherence brings about a local suppression of interference between preferred states selected by the interaction with the environment (Schlosshauer).

Because it is fairly widely accepted that environmental entanglement and the corresponding decoherence processes can play a crucial role in the quantum-to-classical transition and the emergence of classicality from quantum mechanics, the classical behavior of objects tends to be regarded as generated through irreversible interactions with the surroundings (Schlosshauer and Camilleri). Accordingly the origin of local classical properties is thought to originate in the nonlocality of entangled quantum states (Joos, 2002). Decoherence can be seen as a theory of universal entanglement (Joos, website; Zeh, 2006).

A further difficulty with decoherence is that it leaves certain important questions unaddressed or insufficiently addressed (Joos, website). Notably, decoherence does not explain quantum probabilities directly and it does not resolve the quantum measurement problem as it is usually stated, although efforts have been applied in both of these directions (Habib; Joos, website; Zurek, 2009).

However, decoherence itself is an explanation that is inherently complex in principle (Joos, website). Thus, examining simpler approaches would seem to make sense so as to be able to develop as straightforward an understanding as possible of the origin and nature of the quantum-classical transition.

The robustness of classical behavior, both in the aspect that assembling a sufficiently large number of quantum objects together seems invariably to produce a classically behaved object, and for the robustness of classical states with respect to observation or measurement, would also seem to be a reason for seeking further clarification of the quantum-classical transition.

Furthermore, there is evidence that certain additional effects may be involved in quantum to classical transitions (Kofler and Brukner; Leggett, 2005; Adler and Bassi; Bhattacharya et al., 2004; Herzenberg, 2006(a) – 2007(c)). These include, for example, size limitations in measurement processes, continuous spontaneous localization, the effects of continuous observation, and limitations imposed by certain characteristics of our universe as a whole. Thus, there would appear to be a number of different additional physical processes that may be providing mechanisms that could lead to emergent classicality. It would seem that a combination of such different physical effects may together contribute to creating the observed separation or boundary between quantum and classical behavior that characterizes the physical objects in our world.

The borderline between the two different regimes from quantum to classical seems to be by no means precisely identifiable (Bassi). It would seem that we should speak of boundaries rather than boundary, in part, as noted, because different physical processes may be causing transitions between quantum and classical behavior under different conditions. Furthermore, it would seem appropriate also to speak of boundaries because different parts or aspects of a type of physical structure may exhibit quantum-to-classical transitions in the structure's behavior under somewhat different conditions. That is, we should speak of boundaries rather than boundary, insofar as the distinction between classical and quantum behavior may differ not only in different physical systems, but also with respect to the different degrees of freedom or specific types of behavior involved: that is, the boundaries between quantum and classical may differ depending on whether it is the localization of location of an entire object that is in question, or whether conformational changes in shape or changes in rotational or vibrational motion may be under consideration (Bhattacharya et al., 2002, 2004).

To clarify this point with an example, the class of pyramidal molecules exhibit what may be regarded as a particular case of a quantum-classical boundary as a function of their mass and size. This occurs in the behavior of the molecules with respect to inversion. The smallest pyramidal molecule, ammonia, can tunnel quantum mechanically from one state to another via an umbrella inversion (Joos, 2002; Denker). Larger related pyramidal molecules, which do not exhibit such quantum tunneling, tend to maintain a fixed geometrical molecular framework. The distinction in behavior between the ammonia molecule that exhibits quantum mechanical tunneling, and related molecules with similar structures but heavier ligands which do not exhibit quantum mechanical tunneling between the two inversion-related states, can be regarded as a quantum-classical boundary, but is a distinct phenomenon from the quantum-classical boundary with respect to the translational or displacement motion of an entire object that we are primarily considering here.

Identifying classical behavior seems familiar and straightforward, with localization as its most prominent characteristic. Identifying quantum behavior with respect to translational motion seems to be demonstrated most clearly by establishing the presence of interference effects. Quantum behavior has been amply demonstrated for electrons in electron interferometry, and interference experiments have also demonstrated quantum behavior for neutrons, helium atoms, hydrogen molecules, diatomic molecules, and some larger molecules (Arndt and Hornberger). (Experiments on the diffraction of atoms and hydrogen molecules are almost as old as quantum mechanics itself (Leggett, 2002).) Among the largest structures for which quantum behavior has been demonstrated for the entire structure are fullerene molecules; the interference of fullerenes each consisting of 60 or 70 carbon atoms has been demonstrated using interferometry (Arndt and Hornberger; Leggett, 2002). Efforts are under way to attempt to observe interference of even more massive molecules (Arndt and Hornberger). Molecular diffraction experiments on center-of-mass interferences of large molecules such as hemoglobin or even small viruses have been discussed, and further interferometric experiments with even larger molecules will be needed to further explore the characteristics of the behavior of physical structures represented in the size range in the large gap in mass and size between bulk matter and individual atoms or molecules, that is, in the intermediate mesoscopic region in which a gradual transition from well established quantum phenomena to classical appearances takes place (Leggett, 2002).

It would seem to be of interest to continue to examine additional potential candidate processes for causing a quantum-classical transition, in part in order to find out whether simpler explanations than universal entanglement may account for some aspects of the quantum-classical transition which is such a manifest feature of our world.

**Earlier studies: Effects of large-scale properties of the universe**

In the present study, we are interested in following up on earlier studies that have suggested that certain large-scale properties of the universe may have a role in affecting the transition between quantum behavior and classical behavior of objects. These studies appear to have shown that certain properties of the universe related to its expansion or temporal duration that can be expressed in terms of the Hubble constant or similar parameters appear to have a role in effecting the quantum-classical transition.

In the different studies, potential limitations on quantum behavior introduced by different general characteristics of a finite, expanding universe were explored, and estimates were made of limitations on the extent of quantum waves, and of uncertainties in position of objects dependent on general properties of the universe.

These studies have all led to the conclusion that sufficiently large objects must behave classically, and the studies individually introduce threshold criteria for "sufficiently large". Interestingly enough, these earlier papers that were variously based on different physical arguments have come up with roughly similar criteria for a threshold between quantum and classical behavior. Arguments based on the Heisenberg uncertainty principle in the presence of Hubble expansion; random motion in the context of the stochastic quantum theory; wave packet dispersion over time; and constraints on wave packet behavior due to the finite temporal extent of the universe have all led to closely related criteria for a critical threshold size separating quantum behavior from classical behavior (Herzenberg, 2006(a) – 2007(c)).

In the present paper, we show that such criteria can be expressed more directly and simply in terms of a threshold moment of inertia, and we go on to compare this threshold with observed values of the moment of inertia over an extensive range of physical structures.

In the preceding studies, what might be considered non-quantum mechanisms dependent on specific properties of the universe associated with its expansion and finite duration lead, for example, to restrictions on the extent of quantum waves or to quantum uncertainty. These mechanisms appear to lead to the realization of classical behavior for sufficiently large objects, even as the behavior of these objects is described as subject to quantum mechanics. By processes such as the clipping off in time of potentially infinitely extended monochromatic waves into finite wave packets contained within a universe of finite duration, or the introduction of potential uncertainties in velocity and position of semiclassical objects in association with the local expansion of space, physical objects that might otherwise be regarded as completely delocalized quantum objects or completely localized classical objects are subjected to additional localization or delocalization effects affecting the extent of their spatial probability distributions.

Before going on to introduce a new perspective on these results, a brief outline of one approach will be given so as to exemplify where the threshold criteria can come from.

The duration of the universe since its inception at the Big Bang limits the total amount of time available for any physical process, and in particular would seem to set an obligatory maximum temporal extent for any quantum wave. Since the time and frequency spectra of wave functions are related by a Fourier transform (Robinett), this limitation on the temporal extent of quantum waves then essentially causes every quantum wave in effect to become a wave packet, and accordingly leads to an obligatory minimum width in frequency for every quantum wave packet in the universe. As a consequence, there cannot be perfectly monochromatic waves in such a temporally limited universe.

But each such quantum wave is characterized by a wave number/wave length describing its spatial behavior as well as a frequency describing its temporal behavior, and these are related, so that the spatial characteristics of each participating wave must be in accord with its temporal characteristics. As a result, the obligatory minimum width in frequency must be accompanied by an associated minimum width in wave number for such a quantum wave packet. (This obligatory width in wave number will also depend on the object's mass and the wavelength of the quantum wave.) The minimum width in wave number for such a nearly pure frequency wave packet then affects its spatial extent, since spatial wave amplitudes and wave number amplitudes also are related by a Fourier transformation (Robinett). The obligatory spatial width for such a wave packet having a nearly pure frequency and wave number in a temporally limited universe then turns out to be $\Delta x \approx hT/m\lambda$ where T is the duration of the universe, m is the mass of the quantum object, and $\lambda$ is the wavelength of the quantum wave. Thus, no quantum wave packet in the universe can be sharper in frequency than the minimum width in frequency, and these nearly pure frequency waves must also be characterized by this necessary spatial width. Accordingly, even the most nearly monochromatic quantum waves are actually obligatory wave packets and so must have a limiting spatial extent. This limiting spatial extent of wave packets depends inversely on the mass, and so will be smaller for massive objects. Next, the usual specification for minimal wave packets that the spatial width of the wave packet should be comparable to its wavelength is applied (Robinett). This leads to a result that such a quantum wave associated with an object in a temporally limited universe will have a minimum width $\Delta x \approx (hT/m)^{1/2}$. This result could be viewed as a kind of intrinsic core width of the quantum wave structure associated with any object.

The threshold criteria for classical behavior developed in these earlier studies were based on a requirement that the size of the quantum wave structure associated with an object (or the quantum uncertainty in its location) should be smaller in extent than the physical size of the extended object (in contrast to having the quantum wave structure extend beyond the physical size of an extended object, in order to exhibit more typically quantum behavior). Thus, the requirement for more classically behaved objects would be that the region of non-zero probability density associated with the location of the center of mass of the object would be confined to the interior of the object, so that the object would be localized and would not have its wave structure extend appreciably beyond its physical extent; while the requirement for objects to exhibit more typically quantum behavior would be that the quantum probability density would extend beyond the boundaries of the object, so that the object would be more delocalized.

These different approaches have all led to similar criteria for a quantum-classical boundary. In the simplest case, these requirements result in the following equation for an approximate object size separating these two types of behavior:

$$L^2 \approx h/(4\pi m H_o) \qquad (1)$$

Here L is a linear measure of the size of the object; m is the mass of the object; $H_o$ is the Hubble constant, and h is Planck's constant. This equation provides a rough estimate of a threshold size; and the various approaches in the studies referenced above lead to comparable relationships in equations having somewhat different numerical coefficients. For a given mass, objects having sizes considerably larger than this critical value would be expected to exhibit classical behavior in their translational motion, whereas objects having sizes appreciably smaller than this critical value would be expected to be able to exhibit quantum behavior.

The above evaluations were used to define what was identified as the critical size for an object. Thus, the critical size associated with an object of mass m will be given by the equation:

$$L_{cr} = [h/(4\pi m H_o)]^{\frac{1}{2}} \qquad (2)$$

The preceding studies were based on establishing a separation between quantum and classical behavior with respect to translational motion, and discussed the significance of the critical threshold parameters in terms of size, mass, density and related parameters characterizing the object. However, it turns out that the threshold separating classical from quantum behavior based on this requirement can also be expressed and examined even more easily in terms of the magnitude of the moment of inertia of the object in question.

**The moment of inertia and its role**

The moment of inertia of a classical object is given as the sum over the discrete point elements of mass composing the body of the product ($mr^2$), where m is a discrete element of mass and r is the distance of the mass element from a fixed axis; or in the case of a continuous distribution of matter, the moment of inertia is defined as the corresponding integral (Wikipedia, moment of inertia).

We will be concerned primarily with the moment of inertia of an extended object with respect to its center of mass. This can be written in a general form dependent only on the mass and a linear measure of the size of the object. Based on dimensional analysis alone, the moment of inertia I of a non-point object must take the form: $I = kmR^2$, where m is the total mass, R is the radius of the object measured from the center of mass, and k is a dimensionless constant, the inertia constant, that depends on the object's distribution of mass (Wikipedia, moment of inertia). (As a familiar case in point, the moment of inertia of a solid sphere about its center of mass is given by $I = (2/5)mR^2$.)

Thus, it is apparent that, within an order of magnitude or so, the moment of inertia I of a classical object with respect to its center of mass can be estimated at least roughly by the quantity $mL^2$, where L is a length parameter describing the size of the object:

$$I \approx mL^2 \qquad (3)$$

**Recasting the threshold criterion in terms of moment of inertia**

We can write an equation for the critical threshold size of an object that roughly separates quantum behavior from classical behavior in terms of the moment of inertia of the object rather than in terms of its mass and size. If we combine Eqn. (3) with Eqn. (1), we find that the value $I \approx h/4\pi H_o$ gives us an approximate expression for a value of the moment of inertia that would be expected to separate objects that would behave classically from those that would behave quantum mechanically, according to these criteria.

This provides us with an estimate of a threshold value for the moment of inertia that would be expected to separate quantum behavior from classical behavior, which we will designate the threshold moment of inertia:

$$I_{th} = h/(4\pi H_o) \qquad (4)$$

This provides a very straightforward proposed criterion for a boundary separating objects potentially exhibiting quantum behavior from those necessarily exhibiting classical behavior.

Let us put in the numbers. We will use $h = 6.63 \times 10^{-34}$ joule-seconds as the value for Planck's constant, and $H_o = 2.3 \times 10^{-18}$ sec$^{-1}$ as the value for the Hubble constant. Inserting these values into Eqn. (4), we can evaluate a numerical value for the parameter that we have called the threshold moment of inertia in mks units as:

$$I_{th} = 2.3 \times 10^{-17} \text{ kg·m}^2 \qquad (5)$$

This result tells us that, approximately speaking, any object with a moment of inertia larger than about $10^{-17}$ kg·m$^2$ would be expected to behave in a classical manner, while any object with a moment of inertia smaller than about $10^{-17}$ kg·m$^2$ may exhibit quantum behavior. This threshold results from limitations imposed on quantum objects by properties of the universe as a whole that are related to its duration and expansion. This criterion would seem to enable us to set a rough threshold separating generally classical behavior from generally quantum behavior. As noted earlier, other processes can and do contribute to bringing about classical behavior from quantum behavior, and would be expected to lead to classical behavior in various circumstances for objects with moments of inertia below this threshold.

**Comparison of the threshold moment of inertia with some observed moments of inertia of classical and quantum objects**

The moment of inertia is basically a classical concept; however, it has a role in quantum physics also, notably in cases in which the moment of inertia is used in conjunction with the description of rotational states. Moments of inertia of classical objects can be calculated or measured directly; and the differences in energies between the rotational states of a quantum object (such as the band spectra of molecules) can be measured, allowing for an associated moment of inertia to be evaluated from experimental measurements.

To begin with, let's look at the moments of inertia of some large physical objects. For a case of a very large object exhibiting classical behavior we could consider a planet: in the case of the Earth, the moment of inertia of the Earth is approximately $8.0 \times 10^{37}$ kg·m$^2$, some 54 orders of magnitude above the threshold value for a moment of inertia just derived (Wolfram research). As another example of a classical object, an object about a meter in diameter with a density comparable to that of water would have a moment of inertia roughly 50 kg·m$^2$, some 18 orders of magnitude above the threshold moment of inertia. As another example, a small ordinary macroscopic object about a centimeter in size would have a moment of inertia roughly $10^{-7}$ kg·m$^2$, which is about 10 orders of magnitude larger than the critical or threshold moment of inertia evaluated above. Clearly, macroscopic objects including the ordinary ones that we deal with in everyday life, which have masses of grams or kilograms and sizes of centimeters or meters, will have moments of inertia well above the threshold and will easily satisfy this criterion for classical behavior. Accordingly, all very large objects and the objects that we deal with in everyday life must be well within the classical range according to this new criterion.

Next, let's consider slightly smaller objects, for example a spherical object of density 1 gram/cc that has a radius of approximately 0.1 mm. Such an object would have a moment of inertia close to the threshold value for the moment of inertia. Thus, the size of what might be a typical object having a moment of inertia at the critical threshold is somewhat smaller than the sizes of many classical objects, but is still fairly close to the scale of the familiar everyday objects that we recognize as unquestionably classical. This case exemplifies how an object having a moment of inertia approximately equal to the threshold moment of inertia is a small object, somewhat smaller than most familiar human-sized objects, but clearly nowhere near the realm of atomic structure where we are assured that quantum behavior sets in. This leaves a very large range of transitional and quantum behavior at much smaller sizes below the threshold.

We will examine some additional examples of other physical objects, including a range of smaller ones.

To move to a biological example, the largest free-living amoebas grow as large as 1-5 mm (Tan et al.). Such amoebas could have masses larger than a microgram, and moments of inertia in the range of $10^{-12}$ kg·m$^2$. According to the present criterion, these exceptionally large amoebas should exhibit classical behavior.

Bacteria and other prokaryotes typically have sizes of a few microns. Thus they would be expected to have moments of inertia very roughly of the order of magnitude of $10^{-27}$ kg·m$^2$, some ten orders of magnitude smaller than the threshold, and thus these objects would be expected to behave in a quantum manner unless brought into classicality by other effects.

Most viruses that have been studied have diameters between about 10 and 300 nanometers (Wikipedia, Virus). We could estimate that typical viruses might be expected to have moments of inertia of the order of magnitude of $10^{-33}$ kg·m$^2$ to $10^{-35}$ kg·m$^2$; the moment of inertia of a tobacco mosaic virus has been reported in this range (Starodub et al.). Moments of inertia such as these would put viruses well within the range of quantum behavior for entire objects according to the present criterion, and they would be expected to behave in a quantum manner unless brought into classicality by other effects.

In regard to nanoscale structures, in nanotechnology, the generic upper size limit is about 100 nanometers (Goldstein). For a couple of examples: A metal paddle referred to in a numerical study of nanotube-based torsional oscillators is described as having a moment of inertia of $2.4 \times 10^{-38}$ kg·m$^2$ (Xiao and Hou). Also, a micro-motor with equivalent moment of inertia equal to $5 \times 10^{-20}$ kg·m$^2$ is discussed in terms of Newtonian mechanics in an examination of mathematical models and designs for nano- and microelectromechanical systems (Lyshevski). According to the present criterion, these objects would be below the threshold of obligatory classical behavior and could be within the range of possible quantum behavior unless brought into classicality by other effects.

Biomolecular machines may also be of interest (Chowdhury). In regard to macromolecules and polymers, macromolecules range in size roughly between about 1 and 10 nanometers. With respect to polymers, the radius of gyration is directly measurable by light scattering, neutron scattering, and small angle x-ray scattering experiments. Since the radius of gyration $r_g = (I/m)^{1/2}$ this enables evaluation of the moment of inertia (Rudin). A macromolecular assembly such as a ribosome might be expected to have a moments of inertia of the order of magnitude of $10^{-35}$ or $10^{-36}$ kg·m$^2$, in approximate agreement with reported values (Starodub et al.).

A smaller macromolecule, the medium-sized protein lysozyme, is reported to have a moment of inertia of $5 \times 10^{-41}$ kg·m$^2$ (Starodub et al.). It would be well below the threshold of obligatory classical behavior also, and would be expected to behave in a quantum manner unless brought into classicality by other processes.

For present purposes, a very interesting case is that of intermediate size molecules, in particular the fullerenes (Arndt and Hornberger; Krause et al.; Roduner et al.). The diameter of a $C_{60}$ fullerene buckyball is about a nanometer. A value for the moment of inertia of a fullerene buckyball ($C_{60}$) has been referred to in the literature as $1.0 \times 10^{-43}$ kg·m$^2$ (Roduner et al.); additional measurements have been reported for other fullerenes.

Quantum interference experiments with fullerenes ($C_{60}$ and $C_{70}$ molecules) have been carried out (Joos, 2002; Arndt and Hornberger). Research groups have sent fullerene molecules with 60 or 70 carbon atoms each through the equivalent of two-slit interference equipment, dramatically displaying their quantum wave nature as entire objects in translational motion. These quantum interference experiments have established clearly that these fairly large molecules can behave quantum mechanically with respect to their translational motion.

Smaller molecules have even smaller moments of inertia. The moments of inertia for the water molecule with respect to different axes through the center of mass are in the range of about $1 \times 10^{-47}$ kg·m² to $3 \times 10^{-47}$ kg·m² (Chaplin). Rotational transitions of diatomic molecules occur in the microwave region, and can be studied directly using microwave spectroscopy. The moment of inertia for the carbon monoxide molecule is indicated to be $4.49 \times 10^{-49}$ kg·m² (Yates and Johnson). These moments of inertia value are some 31 orders of magnitude below the critical threshold and thus are well into the range of quantum behavior.

Nuclei are even smaller and have smaller moments of inertia: semiclassical spherical rigid-body estimates on the basis of size and mass of nuclei lead to moments of inertia that would range up to about $10^{-53}$ kg·m². As a particular case, the moment of inertia for a rotational transition of the gadolinium isotope $^{152}$Gd has been reported at about $5 \times 10^{-54}$ kg·m² from experimental values of rotational nuclear energy levels (Loveland et al.). Nuclei have moments of inertia in the range of and below roughly 37 orders of magnitude smaller than the critical moment of inertia.

By comparing the magnitudes of these observed moments of inertia with the new threshold criterion, we see that it the magnitudes of moments of inertia of large classically behaved objects are well above the threshold, while the magnitudes of the moments of inertia of physical objects that have been shown by quantum interference experiments to behave quantum mechanically with respect to their translational motion are all far below this new threshold criterion. Thus, the resulting classification would seem to be compatible with the properties of clearly classical and, at least in some instances, clearly quantum objects. In the rather large intermediate mesoscale region, other effects and criteria such as local decoherence can be expected to have a significant role.

**Discussion**

In answer to the question, "Why do macroscopic objects seem to be always well localized in space?", the studies on which this paper is based respond that it is because these objects exist in a finite, expanding universe. We have here recast the quantitative aspect of these studies in terms of the moment of inertias of physical objects.

These effects of a finite, expanding universe on a semiclassical or quantum object might be considered to be due to interaction of the object with its environment in the most general sense, and hence might be regarded as a special case of environmental effects leading to classicality. However, as we have seen, a direct and straightforward rather than general approach has led rather easily to useful results.

This new threshold criterion distinguishes classical from quantum objects on the basis of the magnitude of the moment of inertia of the physical object and proposes a rough boundary between quantum and classical behavior. The resulting classification would seem to be compatible with the properties of clearly classical and, at least in some instances, clearly quantum objects. In the rather large intermediate mesoscale region, and, as noted, other effects and criteria such as local decoherence processes can be expected to have a significant role.

The moment of inertia may be of some value in simply providing a clear and simple measure using a single parameter that ranges over many orders of magnitude between structures that exhibit manifestly quantum behavior, and those that exhibit manifestly classical behavior. The magnitude of the moment of inertia varies rapidly with size and mass, and accordingly could be expected to provide a sensitive parameter that might prove particularly useful for the examination and classification of physical objects, including macromolecular and nanoscale objects, in the large mesoscopic region between well established quantum behavior and manifest classical behavior, where some systems may be neither fully quantum nor fully classical in their dynamics.

The rather interesting question of why the magnitude of the moment of inertia of an object might serve as a parameter useful for classifying and distinguishing objects according to their behavior as predominantly quantum or classical with respect to their translational motion, will be taken up in a separate article.

**Summary**

We have identified a threshold or critical value of the moment of inertia that provides a single parameter separation criterion above which objects should exhibit obligatory classical behavior and below which objects could exhibit quantum behavior, based on certain properties of the universe at large that could potentially affect the quantum-classical transition. This result, that the magnitude of the moment of inertia of an object can be compared to a threshold value requiring classicality, provides a very straightforward approach to the quantum-classical boundary issue. Comparison with observed moments of inertia suggests that this criterion does in fact to provide a reasonable threshold above which classical behavior is always present.

Other processes such as conventional decoherence would seem to be needed in order to understand and predict or retrodict the prevalence of quantum in contrast to classical behavior within the transition mesoscopic region between clearly evident classical behavior and fully quantum behavior.

**Bibliography and References**

Stephen L. Adler and Angelo Bassi, "Is Quantum Theory Exact?" *Science* **325**, 275-276 (2009).

Markus Arndt and Klaus Hornberger, "Quantum interferometry with complex molecules," http://arxiv.org/pdf/0903.1614v1 (9 March 2009), (accessed 19 July 2009).

Bassi, A. and G. C. Ghirardi, *Physics Reports* **379**, 257-426, 2003.

Tanmoy Bhattacharya, Salman Habib, and Kurt Jacobs, "Continuous Quantum Measurement and the Quantum to Classical Transition," LA-UR-02-5289, http://arxiv.org/abs/quant-ph/0211036 (8 November 2002), (accessed 8 August 2009).

Tanmoy Bhattacharya, Salman Habib, and Kurt Jacobs, "The Emergence of Classical Dynamics in a Quantum World, http://arxiv.org/pdf/quant-ph/0407096v1 (14 July 2004), (accessed 1 August 2009).

Martin Chaplin, "Water Structure and Science," http://www.lsbu.ac.uk/water/data.html , (accessed 30 July 2009).

Debashish Chowdhury, "Resource Letter PBM-1: Physics of biomolecular machines," *American Journal of Physics* **77** (7), 583-594, July 2009.

John S. Denker, "The Quantum/Classical Boundary, " http://www.av8n.com/physics/quantum-classical.htm
(accessed 8 July 2009).

Vladimir V. Diky, Natalia V. Martsinovich, and Gennady J. Kabo, Calculation of Pseudorotational Moments of Inertia of Cyclopentane Derivatives Using Molecular Mechanics Method, *J. Phys. Chem.* A, 2001, 105 (20), pp 4969–4973.

Alan H. Goldstein, "Everything you always wanted to know about nanotechnology…," Salon.com, http://dir.salon.com/story/tech/feature/2005/10/20/nanotech/print.html (accessed 31 July 2009).

Salman Habib, "Quantum Classical Transition," http://t8web.lanl.gov/people/salman/deco.html (accessed 27 July 2009).

C. L. Herzenberg, "Becoming Classical: A Possible Cosmological Influence on the Quantum-Classical Transition," *Physics Essays* **19**, 4, 1-4 (2006a).

C. L. Herzenberg, "A possible cosmological effect on the quantum-to-classical transition," http://arxiv.org/abs/physics/0603136 (2006b).

C. L. Herzenberg, "Interpretation of cosmological expansion effects on the quantum-classical transition," arXiv:physics/0606070 http://arxiv.org/abs/physics/0606070 (2006c).

C. L. Herzenberg, "The Role of Hubble Time in the Quantum-Classical Transition," Physics Essays **20**, 1, 142-147 (2007a).

C. L. Herzenberg, "Why our human-sized world behaves classically, not quantum-mechanically: A popular non-technical exposition of a new idea," arXiv:physics/0701155 http://arxiv.org/abs/physics/0701155 (2007b).

C. L. Herzenberg, "The Quantum-Classical Transition and Wave Packet Dispersion," arXiv:0706.1467 http://arxiv.org/abs/0706.1467 (2007c).

Erich Joos, "Decoherence and the transition from quantum physics to classical physics," in *Entangled World: The Fascination of Quantum Information and Computation*, Jürgen Audretsch (Ed.), Wile-VCH Verlag, Weinhem 2002.

Erich Joos, "Decoherence," http://www.decoherence.de/ (accessed 13 July 2009).

Johannes Kofler and Časlav Brukner, "Classical World Arising out of Quantum Physics under the Restriction of Coarse-Grained Measurements," *Phys. Rev. Lett.* 99 , 180403 (2007).

M. Krause, M. Hulman, H. Kuzmany, O. Dubay, G. Kresse, K.Vietze, G. Seifert, C.Wang, and H. Shinohara, Fullerene Quantum Gyroscope, *Physical Review Letters* **93**, 13, 1-4. 24 September 2004.

A J. Leggett, "Testing the limits of quantum mechanics: motivation, state of play, prospects," *Journal of Physics: Condensed Matter* **14**, R415, 2002.

A J. Leggett, "The Quantum Measurement Problem," *Science* **307**, 871, 2005.

Walter D. Loveland, David J. Morrissey, and Glenn T. Seaborg, *Modern Nuclear Chemistry*, Wiley-Interscience, Hoboken, NJ, 2006.

Sergey Edward Lyshevski, *Nano- and microelectromechanical systems: fundamentals of nano- and microengineering*, CRC Press, 2001, p. 63.

Roger Penrose, "On Gravity's Role in Quantum State Reduction," *General Relativity and Gravitation* **28**, 581-599, 1996.

Richard W. Robinett, *Quantum Mechanics: Classical Results, Modern Systems, and Visualized Examples*, Oxford University Press, New York, 1997.

Emil Roduner, , Kosmas Prassides, , Roderick M. Macrae, Ian M. Thomas, Christof Niedermayer, Ulrich Binninger, Christian Bernhard, Anselm Hofer and Ivan D. Reid, "Reorientational dynamics of C60 in the solid state. An avoided level-crossing muon spin resonance study", *Chemical Physics*, Volume 192, Issue 3, 15 March 1995, Pages 231-237.

Alfred Rudin, *The Elements of Polymer Science and Engineering*, Elsevier, 1998.

Maximilian Schlosshauer, "Decoherence, the measurement problem, and interpretations of quantum mechanics," http://arxiv.org/pdf/quant-ph/0312059v4 (28 June 2005), (accessed July 30 2009).

Maximilian Schlosshauer and Kristian Camilleri, The quantum-to-classical transition: Bohr's doctrine of classical concepts, emergent classicality, and decoherence, http://arxiv.org/abs/0804.1609 (10 April 2008), (accessed 10 August 2009).

D. Starodub, R.B. Doak, K. Schmidt, U. Weierstall, J. S. Wu, J. C. Spence, M. Howells, M. Marcus, D. Shapiro, A. Barty, and H. N. Chapman, "Damped and thermal motion of laser-aligned hydrated macromolecule beams for diffraction," *Journal of Chemical Physics* **123**, No. 24, 244304, December 2005.

Olivia Li Ling Tan, Zakaria Ali Moh. Almsherqi and Yuru Deng, "A simple mass culture of the amoeba Chaos carolinense: revisit," *Protistology* **4**(2), 185-190 (2005).

Charles H-T Wang, Robert Bingham and J. Tito Mendonça, "Quantum gravitational decoherence of matter waves," *Class. Quantum Grav.* **23**, L59-L65, 2006.

Wikipedia, "Moment of inertia," http://en.wikipedia.org/wiki/Moment_of_inertia (accessed 17 July 2009)

Wolfram Research, "Moment of Inertia – Earth," http://scienceworld.wolfram.com/physics/MomentofInertiaEarth.html (accessed 30 July 2009).

Shaoping Xiao and Wenyi Hou, Multiscale modeling and simulation of nanotube-based torsional oscillators, *Nanoscale Research Letters*, Volume 2, Number 1 / January, 2007, pp. 54-59.

John T. Yates and J. Karl Johnson, *Molecular physical chemistry for engineers*, University Science Books, 2007.

H. Dieter Zeh, "Roots and Fruits of Decoherence," http://arxiv.org/pdf/quant--ph/05/207802 (8March 2006), (accessed 1 August 2009).

H. D. Zeh, Quantum nonlocality vs. Einstein locality, http://www.rzuser.uni-heidelberg.de/~as3/nonlocality.html (accessed 13 July 2009).

Wojciech H. Zurek, "Decoherence and the Transition from Quantum to Classical – Revisited" *Los Alamos Science*, Number 27, p. 2 – 25 (2002).

Wojciech Hubert Zurek, "Quantum Darwinism," http://arxiv.org/abs/0903.5082 (29 March 2009), (accessed 1 August 2009).

C. L. Herzenberg
carol@herzenberg.net
10 August 2009

qcangmomonlypt1.doc
12 August 2009 draft